\documentclass[smallcondensed,draft,fleqn]{svjour3}
\usepackage{amsfonts}
\usepackage{amsmath}
\usepackage{amssymb}
\usepackage{z-eves}
\usepackage[utf8]{inputenc}
\usepackage{url}
\usepackage{comment}
\usepackage{soul}
\usepackage{xspace}
\usepackage{xcolor}
\usepackage{booktabs}
\usepackage{colortbl}
\usepackage{tabularx}
\usepackage{comment}

\newcommand{\setlog}{$\{log\}$\xspace}

\renewcommand{\Cup}{\mathsf{un}}
\newcommand{\Ris}{\mathsf{ris}}
\newcommand{\In}{\mathbin{\mathsf{in}}}
\newcommand{\Ncup}{\mathsf{nun}}
\newcommand{\Or}{\mathbin{\mathsf{or}}}
\newcommand{\Neq}{\mathbin{\mathsf{neq}}}
\newcommand{\Nin}{\mathbin{\mathsf{nin}}}
\renewcommand{\comp}{\circ}
\newcommand{\Comp}{\mathsf{comp}}
\newcommand{\Dom}{\mathsf{dom}}
\newcommand{\Ran}{\mathsf{ran}}
\newcommand{\Subseteq}{\mathsf{subset}}
\newcommand{\Apply}{\mathsf{apply}}
\newcommand{\Pfun}{\mathsf{pfun}}
\newcommand{\Forall}{\mathsf{foreach}}

\def \dom       {\mathop{\mathrm{dom}}}
\renewcommand{\power}{\mathcal{P}}

\newcommand{\re}{\mathsf{r}}
\newcommand{\w}{\mathsf{w}}
\newcommand{\e}{\mathsf{e}}
\newcommand{\ap}{\mathsf{a}}
\newcommand{\co}{\mathsf{c}}

\title{Automated Proof of Bell-LaPadula Security Properties}
\titlerunning{Automated Proof of Bell-LaPadula Security Properties}

\author{Maximiliano Cristi\'a \and Gianfranco Rossi}

\institute{M. Cristi\'a --- corresponding author \at Universidad Nacional de Rosario and CIFASIS,
Rosario, Argentina -- \email{cristia@cifasis-conicet.gov.ar} \and G. Rossi \at
Universit\`a di Parma, Parma, Italy -- \email{gianfranco.rossi@unipr.it}}

\smartqed

\begin{document}
\maketitle

\begin{abstract}
Almost fifty years ago, D.E. Bell and L. LaPadula published the first formal
model of a secure system, known today as the Bell-LaPadula (BLP) model. BLP is
described as a state machine by means of first-order logic and set theory. The
authors also formalize two state invariants known as security condition and
*-property. Bell and LaPadula prove that all the state transitions preserve
these invariants.

In this paper we present a fully automated proof of the security condition and the *-property for all the model operations. The model and the proofs are coded in the \setlog tool. As far as we know this is the first time such proofs are automated. Besides, we show that the \setlog model is also an executable prototype. Therefore we are providing an automatically verified executable prototype of BLP.

\keywords{Bell-LaPadula model, security, automated proof, \setlog, set theory, binary relations}
\end{abstract}

\section{Introduction}
Computer security is perhaps one of the first application domains where formal methods were thoroughly applied almost from the birth of the field. In fact, computer security was born as part of the defense sector and thus it instantly became a critical application. Nobody could afford the costs of a lethal enemy gaining access to military or intelligence secrets processed and stored in computers. Hence, the software running on those computers ought to be bug-free.

Since the early seventies of the 20th century, the Department of Defense (DoD) of the USA funded R+D projects seeking methods and tools to develop secure software \cite{Anderson00,BLP1,BLP2,DBLP:journals/annals/Lipner15}. In this context, `secure' means software that can keep information confidential. Through the years this problem proved to be (perhaps unexpectedly) subtle and complex \cite{Sabelfeld00}.

Around 1973, D.E. Bell and L. LaPadula, from the Mitre Corporation,
published the first formal model of a secure operating system \cite{BLP1,BLP2}.
Today this model is known as the Bell-LaPadula model, abbreviated as BLP. BLP
is described as a state machine by means of first-order logic and set theory.
They used an \emph{ad hoc} mathematical notation. The model also formalizes two
state invariants known as security condition and *-property. Bell and LaPadula
prove that all the state transitions preserve these invariants.

Over the years, BLP has been thoroughly studied, reviewed and criticized causing a good deal of controversy around it \cite{DBLP:journals/ipl/McLean85,Sabelfeld00}. In a way or another, BLP remains one of the most cited models in the computer security literature and one of a handful secure models that has been implemented and used. In 1982 J. Goguen and J. Meseguer proposed the notion of non-interference as a new definition for secure systems \cite{DBLP:conf/sp/GoguenM82a}. From the dawn of this century the computer security community agrees that non-interference improves on BLP and the former is now considered the dominant approach to the problem of building secure systems \cite{Sabelfeld00,DBLP:conf/esorics/Oheimb04}.

Nevertheless, from a formal verification perspective, in particular  concerning proof automation, BLP is still a challenging problem because of its complexity. Specifically, the automated verification of the security condition and the *-property is a difficult problem. In effect, the formal description of BLP uses several complex set and relational operators (such as those available in formal notations like B \cite{Abrial00} and Z \cite{Spivey00}) and different forms of universally quantified formulas. Actually, both the security condition and the *-property are universally quantified formulas, the latter involving two universally quantified variables. In this regard, we consider BLP as a sort of benchmark in proof automation---in spite of it not being the focus of current research in computer security.

Therefore, in this paper we present a fully automated proof of the security condition and the *-property for all of the BLP model operations. The model and the proofs are encoded in the \setlog (`setlog') tool \cite{setlog}. As far as we know this is the first time these proofs are fully automated and thus the achievement can be regarded as a proof pearl. Besides, due to  properties and features enjoyed by \setlog, the model is also an executable program. Hence, the \setlog model (program) of BLP can be regarded as a correct prototype w.r.t. the security condition and  the *-property.

Formal verification of secure systems has a long and well-established tradition
\cite{DBLP:conf/sp/McLean99}. Interactive theorem provers, such as Coq
\cite{CoqRM} and Isabelle \cite{DBLP:books/sp/NipkowPW02}, have been used to
mechanically verify a range of security problems and systems
\cite{DBLP:journals/jar/BartheBCL19,DBLP:conf/sp/MurrayMBGBSLGK13,DBLP:journals/jcs/BartheGHOB13},
including BLP \cite{CristiaMTh}. Although the expressive power and proving
capabilities of a system such as Coq are incomparable to \setlog's and while
automated proof tactics may help during security proofs, \setlog implementation
of BLP provides a fully automated proof of BLP invariants. This seems not to be
the case with more powerful interactive provers. Furthermore, the fact that the
\setlog model of BLP is also an executable program gets it closer to the
certified programs that can be extracted, for example, from Coq proofs.
However, proposals such as FoCaLiZe \cite{DBLP:conf/pldi/DoligezJR12} use a
combination of manual and automated proofs to prove the correctness of BLP.
FoCaLiZe is an object-oriented programming environment that  combines
specifications,  programs  and  proofs. Proofs in FoCaLiZe are interactive
although the Zenon \cite{DBLP:conf/lpar/BonichonDD07} automatic theorem prover
provides some automation. Differently from \setlog, FoCaLiZe is inspired by
functional programming and type theory. Other works use a variety of formal or
semi-formal verification techniques to analyze properties of access control
models: Stasiak and Zelinski use a model-driven engineering approach to run
simulations on an OCL model of BLP \cite{DBLP:conf/depcos/StasiakZ13}; Haraty
and Naous use Alloy to analyze a role-based access control model
\cite{DBLP:conf/iscc/HaratyN13}; in a work in progress, Devyanin et al. applies
Alloy, Event-B and Rodin to prove properties of a secure operating system
\cite{DBLP:conf/asm/DevyaninKKPS14}. Finally, proof automation does play a key
role in other subfields of computer security such as protocol verification,
mostly by means of different flavors of model-checking
\cite{DBLP:reference/mc/BasinCM18}.

The structure of the paper is the following. In Section \ref{blp} we introduce the BLP model as a summary of the original paper. Section \ref{setlog} presents the \setlog tool showing its constraint solving and proving features. The encoding of the BLP model in \setlog as well as the proof of correctness are explained in Section \ref{setlogblp}. Section \ref{concl} presents our conclusions.

\section{\label{blp}The Bell-LaPadula Model}
In this section we briefly introduce the BLP model and some computer security concepts related to it. We will use the notation of the original report \cite{BLP2}, except when it becomes too obscure. The BLP model is a state machine described by means of first-order logic and set theory.

\subsection{Elements of the model}
After an informal introduction where the problem and the approach to solve it are presented, Bell and LaPadula introduce the elements of the model. The elements are the base sets, state variables and data structures on which the model is built. Some of the key elements are the following:
\begin{itemize}
\item $S = \{S_1,\dots,S_n\}$ is the set of \emph{subjects}. In computer security a subject is any active entity of the system such as a process, a computer, etc.
\item $O = \{O_1,\dots,O_m\}$ is the set of \emph{objects}. In computer security an object is any passive entity of the system such as a file or an I/O device, and also a subject (i.e. $S \subseteq O$).
\item $C = \{C_1,\dots,C_q\}$ is the set of \emph{classifications}, where $C_1 > C_2 > \dots > C_q$. A classification, also called \emph{security level}, indicates the level of access of a subject or the confidentiality level of an object. A classification is called  \emph{security clearance} or just \emph{clearance} when applied to a subject.
\item $K = \{K_1,\dots,K_r\}$ is the set of \emph{categories}. A category, also called \emph{need-to-know}, is a sort of keyword attached to an object or subject.
\item $A = \{\re,\w,\e,\ap,\co\}$ is the set of \emph{access attributes}:
$\re$ead, $\w$rite, $\e$xecute, $\ap$ppend, and $\co$ontrol. $\re$ is
read-only, $\ap$ is write-only, $\w$ is read-write, and $\co$ is the access
attribute that lets a subject modify the access attributes of an object.
\item $F = C^S \times C^O \times (\power K)^S \times (\power K)^O$ is the set of classifications and need-to-know vectors. $\power A$ is the power set of $A$; $B^A$ is the set of all functions from $A$ to $B$. If $(f_1,f_2,f_3,f_4) \in F$ then, $f_1$ is the subject-classification function; $f_2$ is the object-classification function; $f_3$ is the subject-category function; and $f_4$ is the object-category function.
\item $M = \{M_1,\dots,M_c\}$ is the set of access matrices. If $M_k \in M$ then, $M_k$ is a $n \times m$ matrix with entries from $\power A$. The $(i,j)$ entry of $M_k$ shows the access attributes of subject $S_i$ w.r.t. object $O_j$.
\end{itemize}

\begin{definition}[BLP state]
Each state of the BLP model is represented as a triple $(b,m,f) \in \power(S \times O \times A) \times M \times F$ where:
\begin{itemize}
  \item $b$ indicates which subjects have access to what objects in
what mode in a given state;
  \item $m$ indicates the current access matrix;

  \item $f$ indicates the clearance level of all subjects, the classification of all objects and the categories associated with subjects and objects in a given state.
\end{itemize}
$V$ denotes the set of all BLP states.
  \qed
\end{definition}

\subsection{Security condition}
The property called \emph{security condition} is the formalization of the main access rule of DoD's security policy known as \emph{multi-level security} (MLS) \cite{Gasser}. Informally, this access rule can be stated as follows:
\begin{quote}
  A person has the right to \emph{read} a document if and only if the security class of the person dominates the security class of the document.
\end{quote}

A security class is an ordered pair where the first component is a classification and the second component is a set of categories.

\begin{definition}[Security class]\label{d:secclass}
An ordered pair $(n,c)$ is a security class iff $n \in C$ and $c \in \power K$.
\qed
\end{definition}

\begin{definition}[Dominates relation]
Security class $(n_2,c_2)$ dominates security class $(n_1,c_1)$ iff $c_1 \subseteq c_2$ and $n_1 \leq
n_2$. \qed
\end{definition}

With these elements we can state when a state satisfies the security condition.

\begin{definition}[Security condition]
A state $(b,m,(f_1,f_2,f_3,f_4))$ in BLP satisfies the security condition iff:
\begin{gather*}
\forall (s,o,x) \in b: \\
\t1  x = \e \lor x = \ap \lor x = \co
  \lor (x \in \{\re,\w\}
       \land f_2(o) \leq f_1(s) \land f_4(o) \subseteq f_3(s))
\tag*{$\qed$}
\end{gather*}
\end{definition}

\subsection{*-property}
The *-property is not part of the MLS policy. It is necessary only when MLS is implemented on a computer system. This is so because a subject may have read access to an object $o_1$ and write access to an object $o_2$ such that the security class of $o_1$ dominates the security class of $o_2$, i.e. $f_2(o_2) \leq f_2(o_1) \land f_4(o_2) \subseteq f_4(o_1)$. In this case the subject can copy information from $o_1$ into $o_2$ without the control of the security system. Here, Bell and LaPadula assume the standard architecture of a computing system: security is implemented and enforced solely by the operating system; processes perform system calls to gain access to objects; but when they get access, the operating system cannot control the flow of information within each process space. This situation is the essence of the (perhaps unexpected) difficulty of the confidentiality problem in general-purpose computing systems.

Therefore, BLP prevents situations like the one described above by defining model operations that preserve the *-property. In order to state the *-property Bell and LaPadula introduce the following notation:
\[
b(s:x_1,\dots,x_t) = \{o: o \in O \land ((s,o,x_1) \in b \lor \dots \lor (s,o,x_t) \in b)\}
\]
where $b \in \power(S \times O \times A)$, $s \in S$ and $x_1,\dots,x_t \in A$.

\begin{definition}[*-Property]\label{d:starprop}
A state $(b,m,(f_1,f_2,f_3,f_4))$ in BLP verifies the *-property iff:
\begin{gather*}
\forall s \in{}  S: \\
  \t1 b(s:\w,\ap) \neq \emptyset \land b(s:\re,\w) \neq \emptyset \\
  \t1 \implies \forall o_1 \in b(s:\w,\ap); o_2 \in b(s:\re,\w):
    f_2(o_2) \leq f_2(o_1) \land f_4(o_2) \subseteq f_4(o_1)
\tag*{$\qed$}
\end{gather*}
\end{definition}

Informally, the *-property establishes that no subject can access secret objects in read mode if it also has access to `less secret' objects in write mode. Given that the component in charge of enforcing security cannot control what subjects do in their own spaces, the *-property seems a sensible condition.

\subsection{The rules}
BLP defines ten \emph{rules} describing how the model transitions from a state to another. In other words, a rule is a state transition or operation. A rule $\rho$ is a function $R \times V \fun D \times V$, where $R$ and $D$ are model elements which, informally, represent the set of requests ($R$) and the set of decisions ($D$). Hence, a rule receives a request, returns a decision and (possibly) takes the system from a state to another. The ten rules are the following:
\begin{itemize}
  \item \textsf{get-read}, \textsf{get-append}, \textsf{get-execute}, \textsf{get-write}: a subject requests read, append, execute or write access to an object.
  \item \textsf{release-read/write/all/execute}: a subject stops accessing an object in a given mode.
  \item \textsf{give-read/write/all/execute}: a subject grants access to an object to another subject in a given mode.
  \item \textsf{rescind-read/write/all/execute}: a subject cancels the access to an object of another subject in a given mode.
  \item \textsf{change-f}: the $f$ component of the state is changed by a new one.
  \item \textsf{create-object}, \textsf{delete-object}: a subject creates or deletes an object.
\end{itemize}

Next, we show the specification of \textsf{get-write}, which is one of the main rules of the model to give an idea of how Bell and LaPadula used the mathematical language.
\begin{gather*}
\textsf{get-write}: \rho_4((\sigma_1,\gamma,\sigma_2,o,x),(b,m,f)) \equiv\\
\textsf{if } \sigma_1 \neq \phi \lor \gamma \neq \mathsf{get} \lor x \neq \w \lor \sigma_2 = \phi \\
\quad\textsf{then } \rho_4((\sigma_1,\gamma,\sigma_2,o,x),(b,m,f)) = (?,(b,m,f)); \\
\textsf{if } w \notin m(\sigma_2,o) \lor f_2(o) > f_2(\sigma_2) \lor f_4(o) \not\subseteq f_3(\sigma_2) \\
\quad\textsf{then } \rho_4((\sigma_1,\gamma,\sigma_2,o,x),(b,m,f)) = (\mathsf{no},(b,m,f)); \\
\textsf{if } \{q: q \in b(\sigma_2:\re) \land (f_2(q) > f_2(o) \lor f_4(q) \not\subseteq f_4(o)\} \cup{}\\
\quad\{q: q \in b(\sigma_2:\ap) \land (f_2(o) > f_2(q) \lor f_4(o) \not\subseteq f_4(q))\} \cup{}\\
\quad\{q: q \in b(\sigma_2:\w) \land (f_2(o) \neq f_2(q) \lor f_4(o) \neq f_4(q))\} = \emptyset \\
\quad\textsf{then } \rho_4((\sigma_1,\gamma,\sigma_2,o,x),(b,m,f)) = (\mathsf{yes},\mathsf{augb}((\sigma_1,\gamma,\sigma_2,o,x),(b,m,f))); \\
\quad\textsf{else } \rho_4((\sigma_1,\gamma,\sigma_2,o,x),(b,m,f)) = (\mathsf{no},(b,m,f));
\end{gather*}
where
\begin{itemize}
  \item The request is a tuple of five components: $\sigma_i \in S \cup \{\phi\}$, where $\phi \notin S$ is a non-subject element; $\gamma \in \{\mathsf{get},\mathsf{give},\mathsf{release},\mathsf{rescind},\mathsf{change},\mathsf{create},\mathsf{delete}\}$ is a \emph{request element} whose purpose is to indicate the kind of order that must be executed by the rule; $o \in O$; and $x \in A$.
  \item In general, $\sigma_2$ is the requesting subject, $\sigma_1$ is used in some rules as the subject to which a permission is given or rescinded, and $o$ is the object to be accessed.
  \item The component $f$ of the state is assumed to be a tuple of the form $(f_1,f_2,f_3,f_4)$.
  \item $?$, $\mathsf{yes}$ and $\mathsf{no}$ are decisions.
  \item $m(s,o)$ indicates the $(s,o)$ entry of matrix $m$.
  \item $\mathsf{augb}((\sigma_1,\gamma,\sigma_2,o,x),(b,m,f)) = (b \cup \{(\sigma_2,o,x)\},m,f)$.
\end{itemize}

\begin{remark}\label{r:blp}
All the rules have the same interface. The downside of this uniformity is the need to use symbols such as $\phi$ to denote a non-subject and conditions such as $\sigma_1 \neq \phi$ to ensure the rule is called with ``well-typed'' parameters. In modern presentations of MLS only two access modes are considered, read-only and write-only, because all the others can be encoded in terms of these two.

In Section \ref{setlogblp} we will use the \setlog notation to encode BLP. Beyond the syntactic peculiarities of \setlog, our model reflects a modern encoding of BLP, and not a literal translation of it into \setlog. In particular we give each rule an interface reflecting the parameters it needs and nothing else, and we only define two access modes.
\qed
\end{remark}

After giving the specification of a rule the authors prove that it preserves the security condition and the *-property. Formally, they prove the following two lemmas for each rule $\rho$:
\begin{lemma}[Rule $\rho$ is security-preserving]\label{l:secpre}
For any request $r$ and state $v$, if $\rho(r,v) = (d,v')$ and $v$ verifies the security condition, then $v'$ verifies the security condition.
\end{lemma}

\begin{lemma}[Rule $\rho$ is *-property-preserving]\label{l:starpre}
For any request $r$ and state $v$, if $\rho(r,v) = (d,v')$ and $v$ verifies the *-property, then $v'$ verifies the *-property.
\end{lemma}

\section{\label{setlog}The \setlog Constraint Solver}
\setlog is a publicly available satisfiability solver and a set-based,
constraint-based programming language implemented in Prolog \cite{setlog}.
\setlog implements a decision procedure for the theory of \emph{hereditarily
finite sets}, i.e., finitely nested sets that are finite at each level of
nesting \cite{Dovier00}; a decision procedure for a very expressive fragment of
the class of finite set relation algebras
\cite{DBLP:journals/jar/CristiaR20,DBLP:conf/RelMiCS/CristiaR18}; and a
decision procedure for restricted intensional sets (RIS)
\cite{DBLP:conf/cade/CristiaR17,DBLP:journals/corr/abs-1910-09118}. This means
that sets and binary relations are first-class entities of the language. At the
core of these decision procedures is set unification \cite{Dovier03}. The set
terms defined in all these three decision procedures can be combined in several
ways: binary relations are hereditarily finite sets whose elements are ordered
pairs, so set operators can take binary relations as arguments; RIS can be
passed as arguments to set operators and freely combined with extensional sets.
\setlog is an untyped formalism; variables are not declared; typing information
can be encoded by means of constraints. Several in-depth empirical evaluations
provide evidence that \setlog is able to solve non-trivial problems
\cite{DBLP:journals/jar/CristiaR20,DBLP:conf/RelMiCS/CristiaR18,DBLP:conf/cade/CristiaR17,DBLP:journals/corr/abs-1910-09118,CristiaRossiSEFM13}.
Given that \setlog has been extensively described elsewhere, in this section we
will show a few examples for the reader to understand how it works.

In \setlog set operators are encoded as constraints. For example:
$\Cup(A,B,C)$ is a constraint interpreted as $C = A \cup B$. \setlog implements a wide range of set and relational operators covering most of those used in formal notations such as B and Z. For instance, $\In$ is a constraint interpreted as set membership (i.e. $\in$); $=$ is set equality; $\Dom(F,D)$ corresponds to the domain of a binary relation, i.e., $\dom F = D$; $\Subseteq(A,B)$ corresponds to the subset ($\subseteq$) relation; $\Comp(R,S,T)$ is interpreted as $T = R \comp S$ (i.e., relational composition); and $\Apply(F,X,Y)$ is equivalent to $\Pfun(F) \And [X,Y] \In F$, where $\Pfun(F)$ constrains $F$ to be a (partial) function.
Formulas in \setlog are conjunctions (\&) and disjunctions ($\Or$) of constraints; they must finish with a dot (as a Prolog query). Negation in \setlog is introduced by means of so-called \emph{negated constraints}. For example $\Ncup(A,B,C)$ is interpreted as $C \neq A \cup B$ and $\Nin$ corresponds to $\notin$. For formulas to lay inside the decision procedures implemented in \setlog, users must only use this form of negation.

Set terms can be of the following forms:
\begin{itemize}
  \item A variable is a set term; variable names must start with an uppercase letter.
  \item $\{\}$ is the term interpreted as the empty set.
  \item $\{x / A\}$ is called \emph{extensional set} and is interpreted as $\{x\} \cup A$; $A$ must be a set term, $x$ can be any term accepted by \setlog (basically, any Prolog uninterpreted symbol, integers, lists, ordered pairs, etc.).
  \item $\Ris(X \In A, \phi)$ is called \emph{restricted intensional set} (RIS) and is interpreted as $\{x : x \in A \land \phi\}$ where $\phi$ is any \setlog formula; $A$ must be a set term and $X$ is a bound variable local to the RIS. RIS have a more complex structure of which we will show a glimpse in Section \ref{starpropsetlog}; see \cite{DBLP:conf/cade/CristiaR17,DBLP:journals/corr/abs-1910-09118} for a detailed presentation.
\end{itemize}

Being a satisfiability solver, \setlog can be used as an automated theorem prover. To prove that formula $\phi$ is a theorem, \setlog has to be called to prove that $\lnot\phi$ is unsatisfiable.

\begin{example}
We can prove that set union is commutative by asking \setlog to prove the following is unsatisfiable:
\begin{gather*}
\Cup(A,B,C) \And \Cup(B,A,D) \And C \Neq D.
\end{gather*}
As there are no sets satisfying this formula \setlog answers \textsf{no}. Note that the formula can also be written with the $\Ncup$ constraint:
\begin{gather*}
\Cup(A,B,C) \And \Ncup(B,A,C). \tag*{\hfill$\square$}
\end{gather*}
\end{example}

\setlog is also a programming language at the intersection of declarative programming, set programming \cite{DBLP:books/daglib/0067831} and constraint programming. Hence, \setlog programs are basically set formulas.

\begin{example}
If we want a program that updates function $F$ in $X$ with value $Y$ provided $X$ belongs to the domain of $F$ and get an error otherwise, the \setlog code can be the following:
\begin{gather*}
\mathsf{update}(F,X,Y,F\_,Error) \text{ :-} \\
\quad F = \{[X,V]/F1\} \And [X,V] \Nin F1 \And \\
\quad F\_ = \{[X,Y]/F1\} \And \\
\quad Error = ok \\
\quad \Or \\
\quad \Comp(\{[X,X]\},F,\{\}) \And \\
\quad Error = err.
\end{gather*}
Then, \textsf{update} receives $F$, $X$ and $Y$ and returns the modified $F$ in
$F\_$ and the error code in $Error$---think of $F\_$ as the value of $F$ in the
next state. As $\&$ and $\Or$ are logical connectives and $=$ is logical
equality, the order of the `instructions' is irrelevant w.r.t. the functional
result---although it can have an impact on the performance. Variable $F1$ is an
existentially quantified variable representing the `rest' of $F$ with respect
to $[X,V]$. If $[X,V]$ does not belong to $F$ then the unification between $F$
and $\{[X,V]/F1\}$ will fail thus making $\mathsf{update}$ to fail as well.

Now we can call \textsf{update} by providing inputs and waiting for outputs:
\[
\textsf{update}(\{[setlog,5],[hello,earth],[blp,model]\},hello,world,G,E).
\]
returns:
\begin{gather*}
G = \{[hello,world],[setlog,5],[blp,model]\} \\
E = ok \tag*{$\qed$}
\end{gather*}
\end{example}

Since \textsf{update} is also a formula we can prove properties true of it.

\begin{example}
If $Error$ is equal to $err$ then $X$ does not belong to the domain of $F$. In order to prove this we need to call \setlog on its negation:
\[
\textsf{update}(F,X,Y,F\_,err) \And \Dom(F,D) \And X \In D.
\]
Then, \setlog answers \textsf{no} because the formula is unsatisfiable.
\qed
\end{example}

The last \setlog feature we want to show is related to RIS. In fact, the
introduction of RIS in \setlog allows for the definition of \emph{restricted
universal quantifiers} (RUQ). In general, if $A$ is a set, then a RUQ is a
formula of the following form:
\[
\forall x \in A: \phi
\]
It is easy to prove the following:
\begin{equation}\label{e:prop1}
(\forall x \in A: \phi) \iff A \subseteq \{x : x \in A \land \phi\}
\end{equation}
Given that $\{x : x \in A \land \phi\}$ is the interpretation of $\Ris(X \In A,\phi)$, the r.h.s. of \eqref{e:prop1} can be expressed as the \setlog formula:
\[
\Subseteq(A,\Ris(X \In A,\phi))
\]
In \setlog we have defined the $\Forall$ constraint to make RUQ easier to write:
\[
\Forall(X \In A,\phi) \text{ :- } \Subseteq(A,\Ris(X \In A,\phi)).
\]

We use these features to encode BLP in \setlog, to automatically prove Lemmas \ref{l:secpre} and \ref{l:starpre} for each rule, and to provide a correct prototype of BLP in the form of a \setlog program.

\section{\label{setlogblp}Encoding BLP in \setlog}
The \setlog code of the BLP model can be found here \url{https://www.dropbox.com/s/b9hm04kgo3iy6vb/blp.zip?dl=0}. The code includes clauses for each BLP rule; for the security condition and the *-property; and for each proof obligation. Each rule is given by stating its pre- and post-conditions. The clauses encoding proof obligations ensure that all rules preserve the security condition and the *-property as well as some ``typing'' invariants.

The \setlog model is +1 KLOC long excluding comments. It contains 60 proof obligations which take 11.5 seconds of computing time to be discharged\footnote{This number is obtained on a Latitude E7470 (06DC) with a 4 core Intel(R) Core\texttrademark{} i7-6600U CPU at 2.60GHz with 8 Gb of main memory. The software components are the following: Linux Ubuntu 18.04.3 (LTS) 64-bit with kernel 4.15.0-70-generic, and \setlog 4.9.6-18b over SWI-Prolog (multi-threaded, 64 bits, version 7.6.4).}. Around 50\% of this time is consumed in proving that \textsf{getRead} and \textsf{getWrite} preserve the *-property. This is consistent with the fact that these two rules are the most security critical and that the *-property is a rather complex quantified predicate.

In the remainder of this section we will explain the structure of our encoding by presenting the *-property and the \textsf{get-write} operation as representative examples. We will emphasize the modifications we introduced w.r.t. the original model. We close the section showing the structure of the proof obligations.

\subsection{The state of the model}
State variables are gathered in a set named $SState$, for $S$ecurity $State$.

\begin{definition}
Let $Br$, $Bw$, $Fo$, $Fs$ and $M$ be variables; and let $br$, $bw$, $fo$, $fs$ and $m$ be constants. Then, the state of the \setlog model of BLP is defined as follows:
\[
  SState = \{[br,Br],[bw,Bw],[fo,Fo],[fs,Fs],[m,M]\}
\]
where each ordered pair represents one state variable whose meaning is given below. The first component of each pair is a constant that allows easy identification of the corresponding state variable.
\qed
\end{definition}

The two components named $Br$ and $Bw$ correspond to BLP's state component named $b$. In our model we separated the subjects currently accessing objects in read mode ($Br$) from those accessing objects in write mode ($Bw$). Both variables are binary relations whose domain are subjects and whose range are objects.

Connected to the separation of $b$ into $Br$ and $Bw$, is the decision of representing only two access modes: $read$, representing a read-only access; and $write$, representing a write-only access. This is aligned with modern presentations of MLS. Other access modes can be encoded in terms of these. For example, BLP's $\w$ is equivalent to requesting $read$ followed by $write$ (see Section \ref{prototype}).

The next two components of $SState$ correspond to BLP's state component named $f$. However, instead of dividing $f$ into four functions, we divide it into two: $Fo$ is the function associating objects with security classes; and $Fs$ does the same for subjects. Following Definition \ref{d:secclass}, in the \setlog model, a security class is represented as an ordered pair $[L,C]$ where $L$ is a natural number and $C$ is a set. In relation to this we define the following clause:
\[
\mathsf{dominates}([L1,C1],[L2,C2]) \text{ :- }
  \Subseteq(C1,C2) \And L1 =< L2.
\]

The last component in $SState$ corresponds to the state variable with the same name in BLP, that is the access matrix. However, in our model $M$ is a binary relation whose elements are of the form $[O,[S,mode]]$ where $O$ is an object, $S$ is a subject and $mode$ is either $read$ or $write$.

\subsection{\label{starpropsetlog}Encoding the *-property}
Having defined the state variables we encode the *-property in the following clause:
\begin{gather*}
\textsf{starprop}(SState) \text{ :- } \\
\t1 SState = \{[br,Br],[bw,Bw],[fo,Fo],[fs,Fs],[m,M]\} \And \\
\t1 \Forall([S1,O1] \In Br, \\
   \t2 \Forall([S2,O2] \In Bw, \\
          \t4 [Sco1,Sco2], \\
          \t4 S1 \Neq S2 \Or \textsf{dominates}(Sco1,Sco2), \\
          \t4 \Apply(Fo,O1,Sco1) \And \Apply(Fo,O2,Sco2)\\
            \t4 ) \\
       \t3  ).
\end{gather*}
As can be seen, \textsf{starprop} uses two nested $\Forall$ constraints. The innermost $\Forall$ constraint uses some arguments we have not presented so far. These arguments allow for the declaration of existentially quantified variables inside the RUQ ($[Sco1,Sco2]$) which can be used to get the `result' of some (auxiliary) constraints. In this case, we get the security class of $O1$ and $O2$ by means of two $\Apply$ constraints.

The predicate inside the innermost $\Forall$ constraint is logically equivalent to:
\[
S1 = S2 \implies \textsf{dominates}([Lo1,Co1],[Lo2,Co2])
\]
but as \setlog does not provide implication, we encode it as a disjunction.

Now we will argue that our encoding of the *-property is a faithful representation of Definition \ref{d:starprop}. First, note that the antecedent in Definition \ref{d:starprop} is not really necessary as the *-property becomes trivially true when any of those sets are empty. Then, recall that in our model we only have $read$ (read-only, i.e., BLP's $\re$) and $write$ (write-only, i.e., BLP's $\ap$), so $b(s:\w,\ap)$ becomes $b(s:\ap)$ while $b(s:\re,\w)$ becomes $b(s:\re)$. Finally, the quantification over $s$ in Definition \ref{d:starprop} can be reduced to a quantification over those subjects that are accessing some object, because otherwise they pose no risk to the security of the system. Therefore, since in our definition we quantify over $Br$ and $Bw$, we are covering all possibly dangerous subjects.

\subsection{Encoding the rules}
As we have put in advance (cf. Remark \ref{r:blp}), in our model each rule receives the parameters that it needs. For this reason the first \textsf{if} clause of each rule in BLP is not necessary in our model. On the other hand, we have added some pre-conditions that BLP does not include. For example, some of our rules check that an object passed as a parameter belongs to the domain of $Fo$. In our model, \textsf{release-read/write/all/execute} and \textsf{rescind-read/write/all/execute} are divided in two rules: one for releasing/rescinding a $read$ access and the other for $write$ access. \textsf{giveRW} encodes \textsf{give-read} and \textsf{give-write}. Our version of \textsf{change-f} takes as input an object and a security class and replaces the security class of the former with the latter.

Now, as an example, we show our encoding of \textsf{get-write}.
\begin{gather*}
\textsf{getWrite}(SState,S,O,SState\_,Out) \text{ :- } \\
  \t1\textsf{getWriteOk}(SState,S,O,SState\_,Out) \\
  \t1 \Or\
  \textsf{getWriteE1}(SState,S,O,SState\_,Out) \\
  \t1 \Or\
  \textsf{getWriteE2}(SState,S,O,SState\_,Out) \\
  \t1 \Or\
  \textsf{getWriteE3}(SState,S,O,SState\_,Out) \\
  \t1 \Or\
  \textsf{getWriteE4}(SState,S,O,SState\_,Out).
\end{gather*}
where $SState$ is the start state, $S$ is the subject requesting write access to object $O$, $SState\_$ is the after state, and $Out$ is the decision or answer returned by the rule. As can be seen, the specification is divided into five cases. The first one is called the \emph{normal} clause, while the last four are called \emph{abnormal} clauses. The normal clause corresponds to the case when the rule returns \textsf{yes}; the abnormal ones when it returns \textsf{no}.

For example, \textsf{getWriteOk} specifies the case when $S$ can access $O$ in $write$ mode by giving the pre-conditions and post-conditions:
\begin{gather*}
\textsf{getWriteOk}(SState,S,O,SState\_,Out) \text{ :- } \\
\t1  SState = \{[br,Br],[bw,Bw],[fo,Fo],[fs,Fs],[m,M]\} \And \\
\t1  [O,[S,write]] \In M \And \\
\t1  [S,O] \Nin Bw \And \\
\t1  \Apply(Fo,O,Sco) \And \\
\t1  \Forall([Si,Oi] \In Br,[Scoi], \\
\t3        Si \neq S \Or \textsf{dominates}(Scoi,Sco),\Apply(Fo,Oi,Scoi)
         ) \And \\
\t1  Bw\_ = \{[S,O]/Bw\} \And \\
\t1  SState\_ = \{[br,Br],[bw,Bw\_],[fo,Fo],[fs,Fs],[m,M]\} \And \\
\t1  Out = yes.
\end{gather*}
As can be seen, the clause checks that:
\emph{a}) $S$ has $write$ access to $O$ according to the access matrix;
\emph{b}) $S$ is not already accessing $O$ in $write$ mode;
and \emph{c}) all objects accessed by $S$ in $read$ mode have security classes below $O$'s. If all these conditions are met, then $[S,O]$ is added to $Bw$ and the state is updated accordingly.

Finally, the clauses $\textsf{getWriteE*}$ specify the cases where some of the pre-conditions of \textsf{getWriteOk} are not met. In these cases the after state is equal to the start state and the decision is \textsf{no}.

\subsection{Type invariants}
Besides automatically proving that each rule preserves the security condition and the *-property, we have defined and automatically proved the following ``type'' invariants:
\begin{enumerate}
  \item $Fo$ is a partial function; formally: $\Pfun(Fo)$.
  \item $Fs$ is a partial function; formally: $\Pfun(Fs)$.
  \item The range of $Br$ is a subset of the domain of $M$; formally: $\Dom(M,D) \And \Ran(Br,R) \And \Subseteq(R,D)$.
  \item The range of $Bw$ is a subset of the domain of $M$; formally: $\Dom(M,D) \And \Ran(Bw,R) \And \Subseteq(R,D)$.
\end{enumerate}

Verifying that the rules preserve these invariants is important because \setlog is an untyped formalism.

\subsection{\label{proofs}Proof obligations}
All proof obligations are implemented as \setlog clauses. There are six proof obligations for each rule: one stating the invariance of the security condition; one for the *-property; and one for each ``type'' invariant. Each proof obligation is of the form:
  \[
  invariantProperty \And rule \And \lnot invariantProperty\_
  \]
where $invariantProperty\_$ denotes the evaluation of $invariantProperty$ in the after state returned by the rule. This formula is the negation of the standard invariance lemma:
  \[
  invariantProperty \land rule \implies invariantProperty\_
  \]
as \setlog proves unsatisfiability.

The following is the \setlog clause stating that \textsf{getWrite} preserves the *-property.
\begin{gather*}
\textsf{getWritePreservesStarprop}(SState,S,O,SState\_,Out) \text{ :- } \\
\t1  SState = \{[br,Br],[bw,Bw],[fo,Fo],[fs,Fs],[m,M]\} \And \\
\t1  \textsf{starprop}(SState) \And \\
\t1  \textsf{getWrite}(SState,S,O,SState\_,Out) \And \\
\t1  SState\_ = \{[br,Br\_],[bw,Bw\_],[fo,Fo\_],[fs,Fs\_],[m,M\_]\} \And \\
\t1  \textsf{nstarprop}(SState\_).
\end{gather*}

The expected answer for a clause encoding a proof obligation is \textsf{no}, as it corresponds to a formula that is expected to be unsatisfiable.

All the clauses encoding the proof obligations for a particular rule can be called from a clause named $\rho\textsf{Check}$, where $\rho$ is the name of a rule. For example, $\textsf{getWriteCheck}(SState,S,O,SState\_,Out)$ calls (and thus proves) all the proof obligations of \textsf{getWrite}. All proof obligations can be discharged by calling \textsf{checkAll}.

\subsection{\label{prototype}A certified BLP prototype}
Once all the proof obligations have been discharged, we can regard the \setlog model as correct. Given that \setlog models are executable, we can regard the model as a correct prototype. As such, we can run it from different initial states and with different inputs to analyze its behavior from a different perspective. We illustrate this with a simple example.

The following \setlog formula defines some BLP state and checks that the security condition and the *-property are satisfied (this is to ensure the simulation starts from a consistent state).
\begin{gather*}
SS1 = \{[br,Br],[bw,Bw],[fo,Fo],[fs,Fs],[m,M]\} \And \\
Br = \{\} \And Bw = \{\} \And  \\
Fo = \{[o1,[1,\{f14\}]],[o2,[2,\{f14,f15\}]]\} \And \\
Fs = \{[s1,[1,\{cia\}]],[s2,[2,\{f14,cia,f15\}]]\} \And \\
M = \{[o1,[s1,read]],[o1,[s2,write]],[o2,[s2,read]],[o2,[s2,write]]\} \And \\
\textsf{seccond}(SS1) \And \textsf{starprop}(SS1).
\end{gather*}

Now we can run different simulations as the following one:
\begin{gather*}
\textsf{getWrite}(SS1,s2,o2,SS2,Out1) \And \textsf{getRead}(SS2,s2,o2,SS3,Out).
\end{gather*}
where subject $s2$ requests $read$ and $write$ access to object $o2$, in which case the relevant part of the computed answer is:
\begin{gather*}
SS3 = \{[br,\{[s2,o2]\}],[bw,\{[s2,o2]\}], \dots\} \\
Out = yes
\end{gather*}
That is, the model gives $s2$ both accesses to $o2$ because $s2$'s security class is above $o2$'s, and $s2$ is not accessing other objects. A subject calling these two operations on the same object is equivalent to having an operation requesting read-write access to the object.

However, if we run a simulation where $s2$ requests $write$ access to $o1$ and $read$ access to $o2$:
\begin{gather*}
\textsf{getWrite}(SS1,s2,o1,SS2,Out1) \And \textsf{getRead}(SS2,s2,o2,SS3,Out).
\end{gather*}
the answer is:
\begin{gather*}
SS3 = \{[br,\{\}],[bw,\{[s2,o1]\}],\dots\} \\
Out = no
\end{gather*}
That is, the model gives $s2$ access to $o1$ but it does not to $o2$ because otherwise the *-property would be violated as $o2$'s security class is above $o1$'s.

Analyzing a model in this way might save precious time when proofs are run because this tends to decrease the number of failed proof attempts. This is specially appreciated in the context of manual proofs. Tools such as QuickChick \cite{denes2014quickchick} have been proposed along these lines.

\subsection{Further analysis}
So far we have used \setlog to model BLP and to discharge the proof obligations
originally presented by Bell and LaPadula. However, \setlog can be used to
conduct further analysis over the model.

For example, \setlog can be used the check whether or not the sub-clauses
defining each rule form a partition of the input space. In turn, this can be
done in two ways: a) by running simulations representing a partition of the
input space; and b) by proving that the disjunction of the pre-conditions of
each clause is equivalent to $\mathit{true}$ and that these pre-conditions are
pairwise disjoint. Clearly, option b) is better but a) can be used as a first
approximation.

\begin{example}[Analysis by simulation]\label{ex:giverw}
As stated in the BLP model, the \textsf{giveRW} operation does not consider the case when a subject $s1$ tries to
give $read$ access to an object $o1$ to another subject
$s2$ who already has this access. This can be uncovered by
defining a BLP state where $s2$ has $read$ access to $o1$:
\begin{gather*}
\textsf{aState}(SS) \text{ :- } \\
\t1 SS = \{[br,Br],[bw,Bw],[fo,Fo],[fs,Fs],[m,M]\} \And \\
\t1 Br = \dots \And Bw = \dots \And Fo = \dots \And Fs = \dots \And \\
\t1 M = \{[o1,[s1,read]],[o1,[s1,ctrl]],[o1,[s2,read]]\}.
\end{gather*}
and then running a simulation starting from \verb+aState+ where the same access
is given again to \verb+s2+:
\begin{gather*}
\textsf{aState}(SS1) \And \textsf{giveRW}(SS1,s1,s2,o1,read,SS2,Out2).
\end{gather*}
in which case the simulation fails making the error evident.
\end{example}

\begin{example}[Analysis by proof]
The error uncovered in Example \ref{ex:giverw} can be also revealed by attempting to prove that the \emph{negation} of the disjunction of the pre-conditions of each clause of \textsf{giveRW} is unsatisfiable.
\begin{gather*}
(X \Nin \{read,write\} \Or
 [O,[S,X]] \Nin M \Or [O,[S,ctrl]] \Nin M \Or [O,[Si,X]] \In M) \And \\
X \In \{read,write\} \And \\
(X \Nin \{read,write\} \Or [O,[S,X]] \In M) \And \\
[O,[S,ctrl]] \In M
\end{gather*}
If this formula is run on \setlog it answers:
\begin{gather*}
X = read, M = \{[O,[S,ctrl]],[O,[Si,read]],[O,[S,read]]/N1\}
\end{gather*}
which is exactly what was discovered in Example \ref{ex:giverw}.
\end{example}

\section{\label{concl}Concluding Remarks}
We have presented an automated proof of two well-known security properties
carried out with the \setlog tool. These properties are quantified formulas
over the theory of sets and binary relations. \setlog proved to be expressive
enough as to model a complex security model; and proved to be powerful enough
as to automatically discharge, in a rather short time, all the proof
obligations required by the model. Furthermore, as the \setlog model is
executable it can be executed, first, to `test' it and, second, after
discharging the proof obligations, as a correct-by-construction prototype.

As a future work, we plan to develop a tool over \setlog which would
automatically generate the proof obligations needed to verify the invariants of
a state machine. This would not only save precious human time but,
fundamentally, would be less error prone.

\end{document}